# The Distribution and Deposition Algorithm for Multiple Sequences Sets

Kang Ning[1] and Hon Wai Leong[2]

[1]Department of Pathology, University of Michigan, Ann Arbor, MI, USA, 48109 and [2]Department of Computer Science, National University of Singapore, Singapore 117590
kning@umich.edu and leonghw@comp.nus.edu.sg

**Abstract**

"Sequences set" is a mathematical model used in many applications such as scheduling, text process and biological sequences analysis. As the number of the sequences becomes larger, "single" sequence set model is not appropriate for the rapidly increasing problem sizes. For example, more and more text processing applications separate a single big text file into multiple files before processing. For these applications, the underline mathematical model is "multiple sequences sets" (MSS). Though there is increasing use of MSS, there is little research on how to process MSS efficiently. To process multiple sequences sets, sequences are first distributed to different sets, sequences for each set are then processed. Deriving effective algorithm for MSS processing is both interesting and challenging.

In this paper, we tried to formulated the problem of Process of Multiple Sequences Sets (PMSS) by first defined the cost functions and performance ratio. Based on these, the PMSS problem is formulated as to minimize the total cost of process. We have proposed two greedy algorithms for the PMSS problem, which are based on generalization of algorithms for single sequences set. Then based on the analysis of the features of multiple sequences sets, we have proposed the Distribution and Deposition (DDA) algorithm and DDA* algorithm for PMSS problem. In DDA algorithm, the sequences are first distributed to multiple sets according to their alphabet contents; then sequences in each set are processed by deposition algorithm. The DDA* algorithm differs from the DDA algorithm in that the DDA* algorithm distributes sequences by clustering based on a set of sequence features (alphabet content is one of the properties). Experiments show that DDA and DDA* always output results with smaller costs than other algorithms, and DDA* outperforms DDA in most instances. This indicates that distribution of sequences to multiple sets according to sequence features before processing sequences on each set is beneficial. The DDA and DDA* algorithms are also efficient both in time and space.

## 1. Introduction

A sequences set (S) contains one or more sequences, $S=\{s_1, s_2 \ldots s_N\}$. Multiple sequences sets (MSS) contains one or more such sequences sets, $MSS=\{S_1, S_2 \ldots S_M\}$. "Multiple sequences sets" is a mathematical model that has its use in many applications, especially in biological sequences analysis, text process and scheduling.

**Example 1. Multiple oligos arrays**

The synthesis of oligos arrays is important to biological sequences analysis. Since each of the synthesis steps is error prone, and each step is also costly in manufacture process, the oligos have to be deposited onto the array effectively (i.e. short synthesis sequences are required). As the number of oligos used in microarray experiments is increasing, single array is not capable for so many oligos, and these oligos are to be deposited onto different arrays. Therefore, a method to distribute these oligos onto different arrays, and deposit oligos on respective arrays so that the total synthesis cost in minimized is critical.

**Example 2. Text processing (data transfer)**

In data transfer, when single channel is not able to transfer huge amount of data efficiently, multiple channels comes to use. For these multiple channel systems, the appropriate distribution of data to different channels, and effective process to transfer these data though each channel is important.

**Example 3. Scheduling**

In a flexible job shop problem [1], there are many jobs, and each of them has different steps. Usually these jobs can be performed on a single workstation. However, as the scale of this problem becomes so large that one workstation is not suitable, the flexible job shop problem with distributed scheduling on multiple workstations is becoming necessary [1], and the total cost is to be minimized.

These applications are based on the same mathematical model of multiple sequences set. And the underline problem is the process of multiple sequences sets: given a set of sequences and a fixed number $M$, how to process all of the sequences on these $M$ sets efficiently. They all have to accomplish two tasks: distributes sequences to different sets (distribution process), and process the sequences for each set (deposition process). The objective of the whole process is to minimize the total "cost" of processing for all of these sequences on multiple sets (details in "problem formulation"). We refer to these problems as the problem of Process of Multiple Sequences Sets (PMSS).

It is worthwhile to note that the size limitation of single set is not the only reason that multiple sequences set come to use. It will be shown in our analysis that even in situations when all sequences can fit into a single set, switching to multiple sequences model could significantly reduce process cost.

The PMSSS problem is related to the SCS problem, since for each sequences set, the minimum number of process steps is actually the length of the *Shortest Common Supersequence* (SCS) of these sequences.

**SCS problem for single sequences sets**

The problem of finding the *Shortest Common Supersequence* (SCS) of a given set of sequences is a very important problem in sequence analysis. SCS of a set of sequences can be defined as follows: Given two sequences $s = s_1 s_2 ... s_m$ and $t = t_1 t_2 ... t_n$, over an alphabet set $\Sigma = \{\sigma_1, \sigma_2, ..., \sigma_q\}$, we say that $s$ is the *subsequence* of $t$ (and equivalently, $t$ is the *supersequence* of $s$) if for every $s_j$, there is $s_j = t_{i_j}$ for some $1 \leq i_1 < i_2 < ... < i_m \leq n$. Given a finite set of sequences $S = \{s_1, s_2, ..., s_k\}$, a *common supersequence of* set $S$ is a sequence $t$ such that $t$ is a supersequence of *every* sequence $s_j$ ($1 \leq j \leq k$)f in $S$. Then, the *shortest common supersequence* (*SCS*) *of S* is a supersequence of $S$ that has *minimum* length.

The SCS problem has found diverse application in many areas, including data compression [2], scheduling [3], query optimization [4], file comparison [5] and biological sequence analysis [5, 6]. The exact solution for the SCS problem can be computed by dynamic programming. However, the SCS problem on arbitrary number of sequences is NP-hard [7], so many heuristic algorithms have been proposed that give heuristic results for the SCS problem.

A trivial algorithm, called Alphabet [8], gives an approximation ratio of $|\Sigma|$ for SCS problem by using the *periodic supersequence* $S_{ps} = (\alpha_1 \alpha_2 ... \alpha_{|\Sigma|})^K$, where K is the maximum length of sequences, and $\Sigma = \{\alpha_1, \alpha_2, ..., \alpha_{|\Sigma|}\}$ is the alphabets set. However, it does not perform well in practice. Other heuristic algorithms include Majority Merge [7], Tournament [9], Greedy [9], and Reduce-Expand [8]. Several heuristic algorithms were also proposed specifically for computing the SCS of DNA sequences (with $|\Sigma| = 4$). These include Min-Height (MH) [10], Sum-Height (SH) [10] heuristics. (Interestingly, Majority Merge [7] and Sum-Height [10] are the same algorithm.) Recently, we also proposed *look-ahead extensions* of SH heuristic [11], as well as a post-processing reduction procedure (LAP algorithm) [11] for SCS problem. We studied the performance of these algorithms on DNA sequences, and analyzed the use of the above algorithms on the synthesis of oligo array. We have shown that using both look-ahead strategy and post process (both in the LAP algorithm) can generate comparatively short SCS for a set of oligos.

These heuristic strategies for the SCS problem is able to accomplish one of the two tasks for the PMSS problem, that is, process the sequences for each set. The complete sequence of process steps for each sequences set is referred to as the "process sequence" or "synthesis sequence". However, PMSS problem has two tasks, and it is not clear if good heuristic algorithm for SCS problem can be incorporated into the strategy for solving the PMSS problem to achieve good performance.

**PMSS problem**

For the PMSS problem, the objective is to minimize the total "cost" for all sequences sets (details of the cost function defined later). The most important differences between PMSS problem and SCS problem is that for PMSS problem, the sequences are to be distributed to multiple sets before they can be processed on each set. While only deposition process affects the lengths of SCS for a single set of sequences, both distribution and deposition process affect the total "cost" of multiple sequences sets.

An example of multiple sequences sets on $\Sigma = \{A, C, G, T\}$ is illustrated in Figure 1. In this example, we assume there are 12 sequences, and they are two be distributed to 4 sets each consists of 3 sequences. The shortest number of process steps for all of the 12 sequences is ACGACTACTGATG (length of 13), that is the cost of 156 (cost function will be defined later). If we just arbitrarily distribute them to the 4 sets, then the four sets have optimal number of process steps of AGCAGTA, AGACGTAC, AATCGCATC and CGTCATG (length of 7, 9, 8, 7), which are much shorter, and they add up to the much smaller cost of 93. If a distribution method is applied for distribution before process sequences on each set (DDA algorithm), then the total cost can be reduced to 78 (length of 6, 6, 7, 7), and if the DDA* algorithm is applied, then the total cost can be further reduced to 72 (length of 5, 6, 6, 7). (Details of DDA and DDA* are described later.)

| | | | |
|---|---|---|---|
| ACG--T------- <br> --G-C-A-T---- <br> AC-A--A------ | A-C-GT- <br> -GCA-T- <br> A-CA--A | ACAA-- <br> A-A-GT <br> A-A-GT | AC-GT <br> A-AGT <br> A-AGT |
| | **AGCAGTA** | **ACAAGT** | **ACAGT** |
| A--A-----G-T- <br> -----T-C-GA-- <br> AC-C--C------ | AA--G--T- <br> --TCG-A-- <br> A--C-C--C | G--CAT <br> G-AC-T <br> GT-C-T | G--CAT <br> G-AC-T <br> GT-C-T |
| | **AATCGCATC** | **GTACAT** | **GTACAT** |
| A--A-----G-T- <br> --GACT------- <br> AC-AC-------- | A-A-GT-- <br> -GAC-T-- <br> A--C--AC | A-CG-T- <br> -TCGA-- <br> -TC-A-G | ACA--A <br> AC-CC- <br> ACAC-- |
| | **AGACGTAC** | **ATCGATG** | **ACACCA** |
| -C--C-A-T---- <br> --G--T-CT---- <br> -----T-C--A-G | C--CAT- <br> -GTC-T- <br> --TCA-G | -C-C-AT <br> ACAC--- <br> AC-CC-- | TC-G-A- <br> TCAG--- <br> -C--CAT |
| **ACGACTACTGATG** | **CGTCATG** | **ACACCAT** | **TCAGCAT** |

**Figure 1: A simple example of the multiple sequences sets. There are 12 different sequences (first column). Their alignments in 4 different sets (second column), as well as the process sequences (in bold) are given. The different sets and process sequences of DDA algorithm and DDA* (window size of 3) algorithm are also given (third and fourth column).**

There is increasing number of applications on multiple sequences sets. Therefore, it is also important to design efficient algorithms for process of multiple sequences sets to minimize overall cost. However, to our best knowledge, there is few researches [12] on PMSS problem.

In this paper, we have formulated the PMSS problem based on cost functions and performance ratios. Several algorithms have also been proposed for solving the PMSS problem. Two greedy algorithms, Greedy-A and Greedy-D are based on generalization of algorithms for single sequence set, namely, Greedy and SH. We have also proposed the Distribution and Deposition Algorithm (DDA) for PMSS problem. The algorithm first distributes the sequences to multiple sets according to their alphabet contents. Then a deposition algorithm (SH or LAP) is applied to generate process sequence for each individual sets. Another algorithm, the DDA* algorithm, differs from the DDA algorithm in that the DDA* algorithm distributes sequences by clustering based on sequence profiles.

## 2. Computational Model and Algorithms

For multiple sequences sets MSS={$S_1$, $S_2$ … $S_M$}, it is assumed that there are at most *N* sequences in each set S (for simplicity, we assume sequences set are of equal size, $N_i$=N), M is the number of sets, *K* is the maximum length of each sequence, and $q = |\Sigma|$ is the size of the alphabet. There are at most *M*N* sequences altogether. Note that the equal size of the set is just for simplicity, and this setting will not affect the general conclusion. The alphabet content of $\alpha_i$ in a sequence $s_j$ is defined as the number of $\alpha_i$ in $s_j$, divided by the length of $s_j$. In a sequences set *S* with $\Sigma=\{\sigma_1, \sigma_2, …, \sigma_q\}$, the alphabet content of $\alpha_i$ in *S* is defined as the number of $\alpha_i$ in *S*, over the total length of the sequences in *S*.

**PMSS Problem Formulation**

**Max makespan cost**: For each of the sequences sets with $N_i$ sequences, we define the number of process steps $L_i$ as the number of process steps that is needed by the algorithm to completely process the characters. Then we define the *max makespan* cost of each set $S_i$={$s_{i1}$, $s_{i2}$ … $s_{iN}$}, $cost_{MM}(S_i)$, as the multiplication of the number of process steps $L_i$, and the number of sequences $N_i$, $cost_{MM}(S_i)=L_i*N_i$. In other word, for each set, $cost_{MM}(S_i)$ is the multiplication of the max makespan and the number of sequences in $S_i$. The $cost_{MM}$ for MSS is defined as

$$\cos t_{MM}(MSS) = \sum_{i=1}^{M} \cos t_{MM}(S_i) = \sum_{i=1}^{M} L_i * N_i \quad (1)$$

Based on $cost_{MM}$, The PMSS problem can be formally defined as this:

Given a set of sequences, and a constant number $M$, distribute these sequences into multiple sequences sets MSS={$S_1, S_2 ... S_M$}, so that $cost_{MM}$(MSS) is minimum.

For multiple arrays (Example 1), cost function (1) reflects the total cost for mask of size $N_i$ on array $M_i$ with $L_i$ synthesis steps for M arrays. For example, in Figure 1 (third column), $L_1$=6, $L_2$=6, $L_3$=7 and $L_4$=7. The cost $cost_{MM}$ = (6+6+7+7)*3 = 78. However, this cost function is not appropriate for text processing and scheduling problems, for which the sum of completion cost (flow cost) is more meaningful.

**Sum of completion cost**: Given multiple sequences sets MSS={$S_1, S_2 ... S_M$} for $M$ sets, we define completion step $C(s_{ij})$ as the process step at which the algorithm has completed the sequence $s_{ij}$. For example, in Figure 1 (third column), C("ACAA")=4 and C("AAGT")=6. Then we define *sum of completion* cost of each set $S_i$={$s_{i1}, s_{i2} ... s_{iN}$}, $cost_{SC}(S_i)$, as the sum of the completion steps $C(s_{ij})$ for each sequences. $cost_{SC}(S_i)=\sum C(s_{ij})$. Note that for the same algorithm on the same sequence set $S_i$, $C(s_{ij}) \leq L_i$ for every sequences in $S_i$. The cost for MSS is defined as

$$\cos t_{SC}(MSS) = \sum_{i=1}^{M} \cos t_{SC}(S_i) = \sum_{i=1}^{M}\sum_{j=1}^{N} C(s_{i_j}) \quad (2)$$

Based on $cost_{SC}$, The PMSS problem can be formally stated as this:

Given a set of sequences, and a constant number $M$, distribute these sequences into multiple sequences sets MSS={$S_1, S_2 ... S_M$}, so that $cost_{SC}$(MSS) is minimum.

In the following parts, $cost_{MM}$ is used for problems such as multiple array problem, in which max makespan is meaningful. For other problems such as scheduling and text process problems, in which sum of completion is meaningful, $cost_{SC}$ is used.

For one set of sequences $S_i$, the problem of minimizing $cost_{MM}(S_i)$ or $cost_{SC}(S_i)$ is the same problem of finding the *Shortest Common Supersequence* (SCS) of $S_i$={$s_{i1}, s_{i2} ... s_{iN}$}. Since the SCS problem is NP-hard [7], under either $cost_{MM}$ or $cost_{sc}$ cost function definition, the PMSS problem is also NP-hard. Moreover, since PMSS problem is more general than the SCS problem, it is at least as difficult to approximate as the SCS problem.

**Optimal solutions for special MSS**

Optimal solutions are obtainable for special multiple sequences sets (MSS). By "special MSS", we mean that (a) the input sequences are special, and (b) the multiple set parameters (set size, etc.) are also special.

One such special MSS contains very few sequences (no more than 100) in total, as well as very few sequences sets. We denote these as $MSS_{small}$. We also assume that $\Sigma$={0,1}, that is, all of the sequences are binary. For $MSS_{small}$ with M sets each of size N, it is easy to observe that the total number of different multiple sequences sets is $(N*M)!/((N!)^M)$. Therefore, that exhaustive search can be applied on these multiple sequences sets to get optimal results.

**Heuristic algorithms for PMSS problem**

Since PMSS problem is NP-hard, we will turn to heuristic algorithms for solution, except for a few special datasets.

**Alphabet Algorithm**: The simple Alphabet algorithm can be applied on the PMSS problem: for each set, the process sequence is given as a periodical sequence: $(\alpha_1\alpha_2...\alpha_q)^K$. We also call this algorithm "periodic supersequence" algorithm.

**Greedy Algorithms:** Greedy algorithms are based on generalization of greedy algorithms for single sequences set.

We first describe a greedy algorithm based on alignment (Greedy-A). In Greedy-A algorithm, SCS($s_i$, $s_j$) for every pair of sequences are first computed, and each pair(s) of sequences are grouped together. Then these SCS sequences are put back to the sequences pool, and SCS are computed based on these sequences. This process continues iteratively, until the number of sequences in one sequence set becomes N. Then we distribute these sequences on the first set. We continue until all sequences are distributed to M sequences sets. An illustration of this algorithm is given in Figure 2 (a). Note that the Greedy-A algorithm is based on sequence-by-sequence approach for sequence distribution, which inherently make it very slow and not scalable to many sequences.

Greedy-A has $O(K^2N^2M^2)$ time complexity and $O(KNM+N^2M^2)$ space complexity. Unfortunately, it was shown in [11] that Greedy-A do not necessarily has better cost than Alphabet algorithm.

Another greedy algorithm is based on deposition (Greedy-D). This algorithm first process all of the sequences together using SH algorithm [10]. When there are N sequences completed, they are outputted onto one set, and the remaining sequences are re-processed. This process continues until all of the sequences are distributed. An illustration of this algorithm is given in Figure 2 (b). This algorithm is based on chracer-by-chracter approach, which is very fast and quite scalable.

Greedy-D has $O(KNM^2)$ time complexity and $O(KNM)$ space complexity.

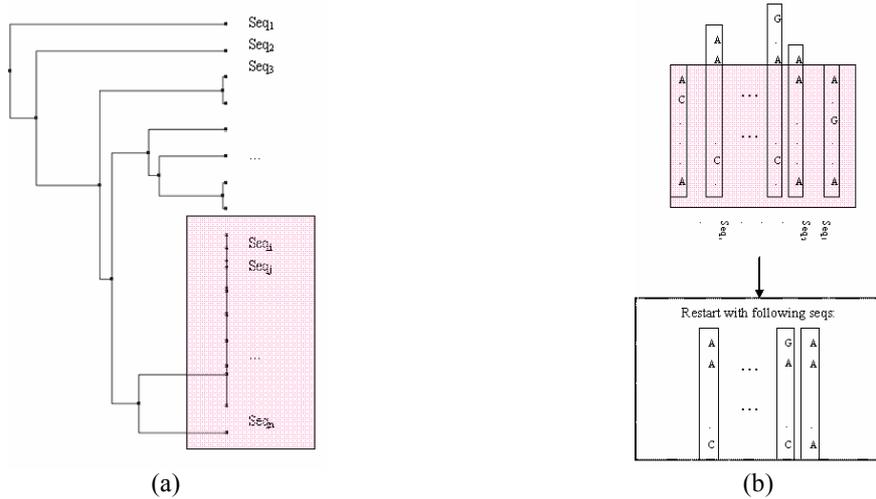

(a)            (b)

**Figure 2. The illustrated process of (a) Greedy-A, (b) Greedy-D. The shadowed oval and box indicate the sequences sets. In (a), each leaf represents a sequence, in (b), each vertical bar represent a sequence.**

**Distribution and Deposition Algorithm (DDA) and DDA\* algorithm:** In this section, we proposed two algorithms, the DDA algorithm and the DDA\* algorithm. The DDA algorithm composes of two parts: the distribution process distributes sequences to different sets, and deposition process generates short process sequence for each set. The initial idea of DDA algorithm is already introduced in [12], but since the idea is important, we present the general framework as below. The DDA\* algorithm differs from the DDA algorithm in that the DDA\* algorithm distributes sequences by clustering technique based on sequence profile.

- **Distribution Method for the DDA algorithm**

For the PMSS problem, the cost of each set is proportional to the number of process steps, and the number of process steps is affected by the alphabet contents of each sequences set.

In an experiment descried in previous paper [12] based on $\sum=\{A,C,G,T\}$, "GC content" is defined as the sum of alphabet contents of "G" and "C". The process steps are generated by LAP algorithm [11]. Results show that the numbers of process steps are smaller for sequences sets with GC content of 20% and 50%, compared to sets with GC content 30% and 40%.

Based on these observations, it is clear that in the distribution process, the alphabet content should better be chosen so that (a) sequences in the same set have similar alphabet contents and (b) the average alphabet content for a specific alphabet $\alpha_i$ should be either high (i.e. 80% GC contents) or low for a single set. Therefore, we have designed the distribution method which first sorts sequences according to alphabet contents; there is a sorted list for each of the alphabet. Then sequences are selected from these sorted lists and put to multiple sets in a round robin manner so that on a single set, one of the alphabets $\alpha_i$ has much higher contents than other alphabets. Sequences that do not have biased alphabet content are distributed to sequences set that have most similar alphabet contents.

Take DNA oligo arrays, in which the "GC content" is considered, for example. Figure 3 shows how distribution process is applies so that both criteria (a) and (b) are taken into consideration. Note that for arrays with GC contents around 20% and 80%, the GC contents of individual sequences are also similar on the same array; but for arrays with average GC contents near 50%, criteria (b) could only be partially fulfilled.

|  | GC content (↓) |  |
|---|---|---|
| $Seq_1$ | 0.20 | |
| $Seq_2$ | 0.20 | Cluster 1 |
| $Seq_3$ | 0.21 | |
| … | … | |
| … | … | … |
| … | … | |
| $Seq_{i-1}$ | 0.50 | Cluster i |
| $Seq_i$ | 0.51 | |
| … | … | |
| … | … | … |
| … | … | |
| $Seq_{M*N-2}$ | 0.88 | Cluster M |
| $Seq_{M*N-1}$ | 0.89 | |
| $Seq_{M*N}$ | 0.89 | |

**Figure 3. An example of distributing oligos to different arrays. The sequences are sorted by GC contents from left to right. Oligos in array1 (red) and array M (blue) are of high and low GC contents, respectively. The oligos in array *i* (green) are of average GC contents of 50%.**

This distribution method has time complexity of O(*KNMq*log(*NM*)) [12].

- **Distribution Method for the DDA\* algorithm**
  To make the distribution process more effective, we have proposed the DDA\* algorithm that is an improved version of the DDA algorithm. The DDA\* algorithm differs from the DDA algorithm in that while DDA algorithm simply distribute sequences according to their alphabet contents, the DDA\* algorithm distributes sequences by clustering technique based on sequence profile (in which alphabet content is one feature).

**Features**
For each sequence, the sequence profile is composed of a set of sequence features. The features that we have used in DDA\* are the alphabet contents of motifs in each sequence *s*. Each motif is a short substring of the original sequence *s*. We have adopted a sliding window technique to get motifs. Based on empirical results (details not shown), we have set the sliding window size as w=|∑|, and obtained the K-w+1 motifs starting at position 0, 1, … K-w in each sequence. The reason that w=|∑| can yield good results is probably that w=|∑| is the minimum size of motif so that the abundance of different alphabets in the motif can be distinguished. For each of these motifs, the alphabet contents for each of the alphabet $α_i$ is computed, and then treated as a feature for sequence *s*. By this method, there are (K-|∑|+1)*|∑| features for *s* in total, which is proportional to the multiply of length of *s* and alphabet size |∑|. Figure 4 gives an example of generating features based from sequences.

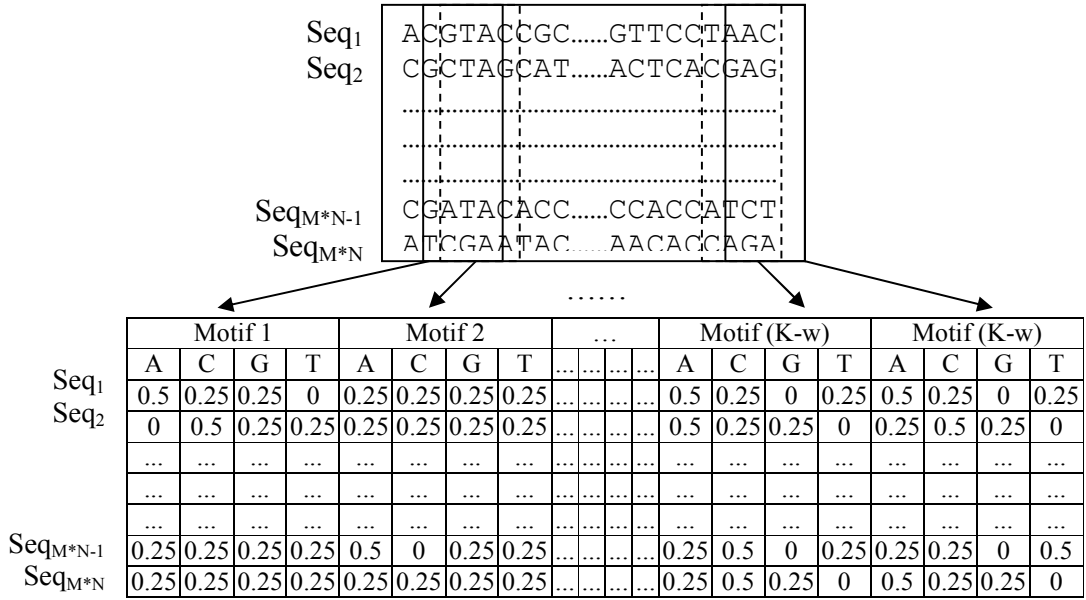

**Figure 4. An illustrated example of generating features based on sliding window from sequences.**

These features are chosen because these features (the alphabet contents of motifs) reflect the probabilities that certain alphabets at different positions in multiple sequences can be processed in a single step. For different sequences, similar features indicate that they are more likely to have similar process steps.

**Clustering**

We have used the Expected Maximization (EM) clustering algorithm for clustering sequences based on their features. The number of clusters is set to be the number of sequences sets (M). After clustering by EM, we have applied post process to adjust the number of sequences in each set.

To make the whole process efficient, the post process is designed to be a simple heuristic method. First, for those clusters (sequences sets) having more than $N$ members, we select those sequences which are outliers. Here we define an "outlier" to be a sequence $s'$ such that if $s'$ is removed from the set, then the process cost for the remaining sequences can be reduced. For example, in Figure 1 (first column), when we use the set size of 2, then "ACAA" will be the outlier for the first set. These outliner sequences form an "outliner sequences pool". Since 1) deposition process of multiple sequences is efficient by SH or LAP algorithm (details below), and 2) the outliers can be selected in batch, the outlier selection could be very efficient. Then for each of the clusters that has less than $N$ members, we add in a sequence, whose feature is most similar to the members of this cluster, from "outliner sequences pool" one by one. Here the "similar" sequence $s''$ is defined such that the addition of $s''$ to the set will increase the process cost minimally. Addition of sequences to clusters could also be efficient in batch process.

Note that there are two extremes of this distribution process. One is that the window size is equal to the length of the sequence ($w=K$), so that only the alphabet contents of the whole sequences are used as features. Based on this sliding window, the distribution method is similar to DDA's distribution method. Another is that the window size is equal to the one character ($w=1$). In this situation, the distribution method is similar to the character-by-character strategies used by Greedy-D.

- **Deposition Method**

Sum Height (SH) algorithm [10] for single sequences set (S) is a simple while efficient deposition method. Suppose we analyze every sequence in a batch from left to right, the *frontier* is defined as the right most characters to be analyzed. Initially, the process sequence $CS$ is empty for set $S$. At each step, let $t$ be the majority character among the "*frontier*" characters of the remaining portions of the sequences in set $S$. Here we break the tie arbitrarily. Set $CS=CS||t$ (where "||" represent concatenation) and delete the "*frontier*" $t$ characters from the corresponding sequences in $S$. Repeat until no sequence is left.

A natural way to improve the SH algorithm is to apply a "*look-ahead*" strategy to it [10]. This strategy looks at a number of steps ahead before deciding which character(s) is the *best* to be added. More specifically, choose two integers *m* and *l* such that *l*≤*m*. Then, the *look-ahead extension of the SH method* works as follows: (i) Examine all possible partial processed sequences that can be generated in *m* steps; (ii) for each such generated partial process sequences, compute the resulting sum height; and (iii) select the partial process sequence that will result in the *largest increase of sum-height in m steps*, and extend the chosen process sequence by *l* (≤*m*) characters. (It can be shown that extending by *l* characters (instead of *m*) gives the potential of obtaining even better increase in the sum-height after *m* steps.) Here again, we break the tie arbitrarily. This look ahead algorithm is called (*m*, *l*)-look-ahead SH, and is abbreviated to (*m*, *l*)-*LA-SH*. Previous research[11] have shown that (*3, 1*)-*LA-SH* has best balance between the process sequence lengths and process time.

The LAP algorithm is a post process algorithm that first generate a process sequence (template) by SH or (*m, l*)-*LA-SH* algorithms for all of the sequences in a set and then uses post-process to further reduce the length of this template. The method achieve this by iteratively check every character in the template. If after removing a certain character, the remaining sequence is still the common supersequence, then a shorter template is obtained. The method continues until there is no character removable. For example, given three sequences "ACGT", "CGGT" and "AGCT", suppose the template is "ACGGTTC". Then the application of LAP algorithm can reduce this template to "ACGGTC" by removing "T".

For the LAP algorithm on a single set, the time complexity of (3,1)-LA-SH is $O(N^3K)$, and the time complexity of the post process is $O(N^3K^2)$. Therefore, the time complexity of the LAP algorithm based on (3,1)-LA-SH for M sequences sets is $O(MN^3K^2)$.

- **DDA and DDA* algorithms**

We refer to DDA algorithm based on SH as DDA-SH, and DDA algorithm based on LAP as DDA-LAP. Overall, the DDA-SH algorithm has time complexity of $O(MNK + KNMq\log(NM))$, and the DDA-LAP algorithm has time complexity of $O(MN^3K^2 + NMq\log(NM))$. An example result of DDA algorithm (DDA-LAP) on PMSS problem is shown in Figure 1 (column 3).

Similarly, the DDA* algorithm based on SH is referred to as the DDA*-SH algorithm, and the DDA* algorithm based on LAP is referred to as DDA*-LAP algorithm. The distribution method for the DDA* algorithm is dependent on the clustering technique, which is highly dependent on the number of iterations. However, since current clustering algorithms are quite fast, both DDA*-SH and DDA*-LAP algorithms are fast in practice (details in Experiments). An example of DDA* algorithm (DDA*-LAP) on PMSS problem is shown in Figure 1 (column 4).

**Performance Ratios and Lower Bounds**

For special multiple sequences sets as mentioned above, the optimal process sequences can be computed. However, for arbitrary multiple sequences sets, optimal results are not obtainable. Therefore, we have computed the *performance ratios* as indicators of the performance of different algorithms.

For each sequences set, the simplest *periodic supersequence* by Alphabet algorithm $S_{ps}=(\alpha_1\alpha_2\ldots\alpha_{|\Sigma|})^K$ has length of qK (q=|Σ|). Thus for multiple sets MSS={$S_1, S_2 \ldots S_M$} with N sequences per set, and each of length K, the periodic supersequence gives the "max makespan" cost of qKMN. Here we use periodic supersequence as a standard, and divide $cost_{MM}(MSS)$ of an algorithm by qKMN to generate performance ratios. For simplicity, the $cost_{SC}(MSS)$ by an algorithm are also divided by qKMN to generate performance ratios based on "sum of completion". The *performance ratio* of the algorithm *A*, $R_A(MSS)$, is defined as

$$R_A(MSS)= cost_A(MSS) / (|S_{ps}|*MN) = cost_A(MSS) / (qKMN) \qquad (3)$$

In which $cost_A(MSS)$ is the cost ($cost_{MM}(MMS)$ or $cost_{SC}(MSS)$) for MSS by algorithm *A*. Smaller performance ratios indicate better solution. Notice that this "performance ratio" definition is different from the usual notion of the performance ratio of an approximation algorithm [13]. The reason is that not only the optimal result for PMSS problem is not available, but there is also no good bound for optimal result of PMSS problem. But it has to be emphasized that the ratio $R_A(MSS)$ is a normalized score that can be used to compare the performance of different algorithms.

It is obvious that Greedy-D ≤ 1 since the synthesis sequences produced by SH is bounded by qK. Moreover, $R_{DDA}(MSS) \leq 1$ and $R_{DDA*}(MSS) \leq 1$ since the synthesis sequences produced by LAP is also bounded by qK.

We have also computed the *lower bound* for SCS of all sequences in MSS (abbreviated as lower bound), as another indicator to evaluate different heuristic algorithms on different settings. To compute this lower bound, we first selected $q$ ($q=|\Sigma|$) sequences from all of the sequences; each of them contains the largest number of alphabet $\alpha_i \in \Sigma$. Then, we apply dynamic programming on these sequences to get optimal number of process steps for these $q$ sequences. Such number is then multiplied by the number of sets (M), which provides us the "lower bound". If $q$ is too large for dynamic programming, we just select fewer sequences so that 1) dynamic programming can be applied, and 2) the lower bound thus computed is maximized. Note that since this lower bound is computed based on all of the sequences, it is not necessarily smaller than the cost of DDA and DDA* algorithms in theory. However, it is a standard against which the performance of different algorithms can be compared. And in practice, this bound is much smaller than the cost of DDA and DDA* algorithms.

## 3. Experiments

**Datasets and Experiment Settings**
The DDA and DDA* programs are implemented in C++ and Perl. The experiments are performed on a PC with 3.0GHz CPU and 1.0GB RAM, running on a Linux system.

For simulated sequences sets, we have randomly generated sequences with specific length and alphabet content. The analysis on real sequences is more important. We have selected real sequences from DNA oligos arrays, as well as real text sequences. For real oligo sequences, we have randomly chosen gene sequences from EST sequences, plant sequences and rodent sequences datasets obtained from the well-known GenBank [14] (http://www.ncbi.nlm.nih.gov/genbank). To approximate real applications, we have chosen the Primer3 software [15] to select good probes (oligos) from genes. By using the Primer3 software, the probe length, extend of complimentarity and temperature (default values) are thus optimized. After suitable probes are selected from each of the genes, we check all of the probes, and remove those which will cause cross-hybridization.

The real text sequences are obtained from 20 Newsgroup DataSet at CMU Text Learning Group Data Archives (http://www.cs.cmu.edu/afs/cs.cmu.edu/project/theo-20/www/data/news20.html). The multiple sequences sets are useful here because there is a huge number (19,997) of text sequences in the datasets to be processed. We have separated the sequences to 5 groups, they are "comp" dataset (5000 seqs), "rec" dataset (4000 seqs), "sci" dataset (4000 seqs), "talk" dataset (4000 seqs) and "misc" dataset (2997 seqs), corresponding to different contents.

We have first accessed the performance of distribution method and deposition method (of DDA and DDA* algorithms) separately. Then we have compared different heuristic results agasint optimal results on special MSS. After these analyses, we have analyzed the performance of DDA and DDA* algorithm, and compared them with greedy algorithms on simulated sequences and real sequences. We emphasis that one of the comparisons is to compare DDA and DDA* on multiple sequences sets with deposition algorithms on a single set (i.e., a single set that containing all of the sequences in MSS). The purpose of this comparison is to show that even there can be very large set so that all sequences can fit in, the cost of DDA and DDA* on multiple sequences sets is still better.

In the following parts, the cost function $cost_{MM}$ is used for multiple array problems, the $cost_{SC}$ cost function is used for scheduling and text processing problems; and both of them are used for comparison of heuristic results against optimal results.

**Results**
- **Assessment of the Distribution Method and Deposition Method**
In this part, we have assessed the distribution method and deposition method separately.

For the DDA algorithm, we have calculated the alphabet contents of different sequences sets. We have used DNA sequences based on $\Sigma=\{A,C,G,T\}$ for analysis. It is easy to see that sequences based on other alphabets have similar results. The general results are similar to those in our previous paper [12]: (a) the GC contents of the sequences on each set are very similar and (b) most of these sets have average GC contents of below 20%, above 80%, or around 50%. Since similar results can be observed on almost all datasets, the distribution process is effective. Results on text sequences (details not shown here) also show that the distribution process can distribute sequences so that alphabet contents are well distributed.

Then we have assessed the distribution method for DDA* algorithm. Since the clustering technique is used for the distribution process in DDA*, we have analyzed the "log likelihood" of the resulting clusters.

For each set, small "log likelihood" indicates the similarity of the sequence features. Results show that most of the distribution results have "log likelihood" less than -1, and more than half of the datasets examined result in "log likelihood" less than -2. Since we have previously explained that the similar features of sequences indicate that they are more likely to have similar process steps, these small "log likelihood" indicates small process cost. In post process to adjust the set size, the approach that we have adopted is also beneficial for small process cost. Therefore, the distribution method for DDA* algorithm is effective.

The effectiveness of the SH and LAP deposition algorithms has been empirically proven by our previous research [12]: for a set of sequences with arbitrary alphabets, the LAP deposition algorithm can generate smaller number of process steps than Alphabet [8], Majority Merge [7] (Sum-Height [10]), Tournament [9], Greedy [9] and Reduce-Expand [8] algorithms.

- **Optimal and Heuristic Results on Special Datasets**

To assess the performance of heuristic algorithms, we have compared optimal results ($R_{Opt}$) with DDA-SH ($R_{DDA-SH}$(MSS)), DDA-LAP ($R_{DDA-LAP}$(MSS)), DDA*-SH ($R_{DDA*-SH}$(MSS)) and DDA*-LAP ($R_{DDA*-SH}$(MSS)) on special MSS. These special MSS include three randomly generated $MSS_{small}$ (small_1, small_2 and small_3). We have randomly generated 10 datasets for three different settings, and the average results are given. The performance ratio based on $cost_{MM}$ and $cost_{SC}$ are given in Table 1.

**Table 1. Performance ratio of heuristic algorithms based on $cost_{MM}$ / $cost_{SC}$, compared with optimal results.**

| Dataset | M*N | N | K | $R_{Opt}$ | $R_{DDA-SH}$ | $R_{DDA-LAP}$ | $R_{DDA*-SH}$ | $R_{DDA*-LAP}$ |
|---|---|---|---|---|---|---|---|---|
| Small_1 | 10 | 5 | 10 | 0.750 / 0.704 | 0.806 / 0.755 | 0.789 / 0.741 | 0.806 / 0.755 | 0.789 / 0.741 |
| Small_2 | 20 | 10 | 10 | 0.836 / 0.821 | 0.900 / 0.878 | 0.887 / 0.852 | 0.875 / 0.838 | 0.850 / 0.835 |
| Small_3 | 50 | 10 | 5 | 0.800 / 0.765 | 0.865 / 0.836 | 0.844 / 0.823 | 0.820 / 0.779 | 0.800 / 0.770 |

Based on the results of Table 1, we observe that though the performance ratios of DDA and DDA* ($R_{DDA-SH}$, $R_{DDA-LAP}$, $R_{DDA*-SH}$ and $R_{DDA*-LAP}$) are larger than $R_{opt}$; for small datasets, the gaps are small. For example, the results of DDA*-LAP and optimal results on Small_3 only differ slightly. These indicate the superior performance of DDA and DDA* algorithm. DDA-LAP outperforms DDA-SH a little bit. And DDA* algorithm performs better than DDA algorithm.

- **Results on Simulated Sequences**

For simulated sequences, we have used $\sum$={A,C,G,T} (same as in DNA sequences), the lengths of the sequences K = 25 and 35 (longer sequences have similar results), and the size of the sets N = 10,000 and 20,000 (other set sizes have similar performance to 20,000, refer to [15]). For each of the settings (N, M and K), we have randomly generated 10 datasets each of N*M sequences. The results are based on the average number of process steps for each setting. The $cost_{MM}$ cost function is used for analysis.

We have compared DDA with Greedy algorithms. To analyze the effect of using multiple sets against single set, the results of SH ($R_{SH}$) and LAP($R_{LAP}$) on a single set *S* (that is, application of SH and LAP on a set with *all of the sequences* in *MSS*) are computed, and compared with the results of DDA and greedy algorithms on MSS. The results of lower bounds ($R_{LB}$) are also computed. Results are shown in Figure 5.

**Figure 5. Performance ratios of DDA, DDA* and other algorithms on simulated datasets. X-axis shows the values of K, M*N and N in vertical order. The three cases have different number (M*N) of total sequences.**

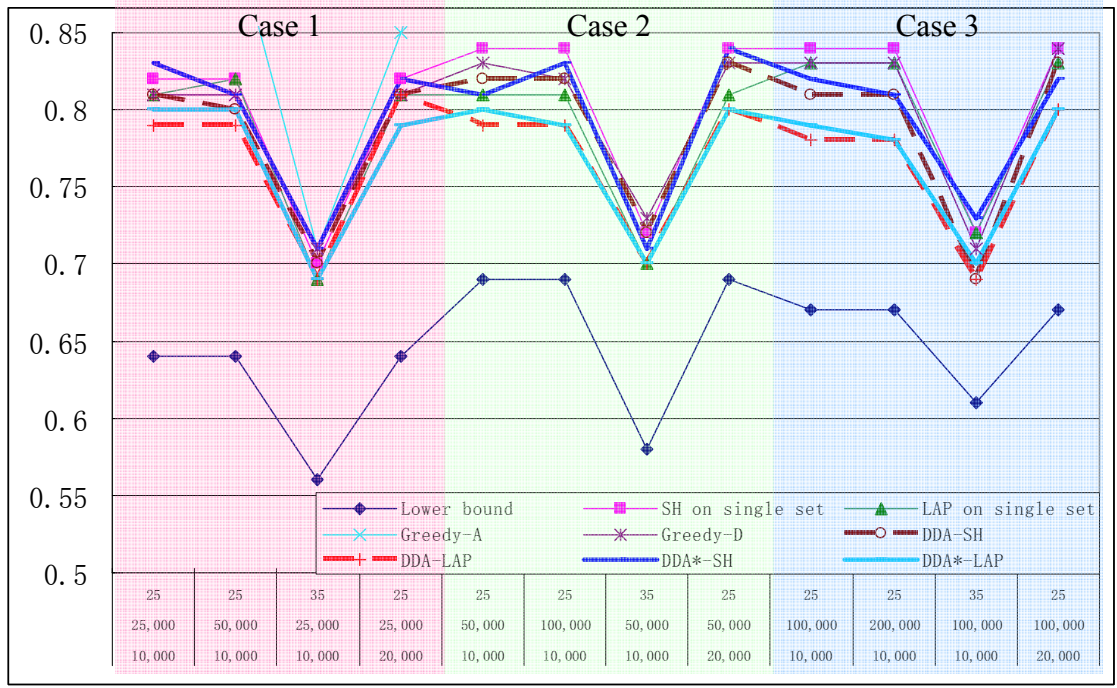

From the result in Figure 5, we can see that except for Greedy-A, all of the performance ratios are smaller than 0.85, much better than $R_{Alphabet}$ (performance ratio of 1.0). $R_{Greedy-A}$ is larger than 1.00 on many datasets. The time needed by Greedy-A is also magnitudes larger than other algorithms. So this greedy algorithm has bad performance, and will not be compared with other algorithms in later parts. $R_{Greedy-D}$ is smaller than 1, but still a little bit larger than $R_{DDA-SH}$, $R_{DDA-LAP}$, $R_{DDA*-SH}$ and $R_{DDA*-LAP}$. Compared between results on multiple sets and those on single set, we observe that the performance ratios $R_{DDA-SH}(MSS)$ and $R_{DDA-LAP}(MSS)$ are constantly better than the results of SH and LAP on single sequences set. This is significant when the number of sequences is larger ($M*N \geq 100,000$). We also observed that $R_{Greedy-D}(MSS)$ is constantly larger than $R_{DDA-SH}(MSS)$ and $R_{DDA-LAP}(MSS)$, while $R_{DDA*-SH}$ and $R_{DDA*-LAP}$ have the best results. This indicates that distribution of sequences by alphabet contents significantly affect performance ratios. Additionally, performance ratios of different algorithms on smaller sets (N=10,000) are constantly smaller than those on sets with N=20,000. It is also observed that performance ratios on longer (35) sequences are smaller than those on shorter (25) sequences. The lower bounds of the performance ratios are about 0.6 to 0.7, which are about 0.1 to 0.2 less than $R_{Greedy-A}$, $R_{DDA-SH}$, $R_{DDA-LAP}$, $R_{DDA*-SH}$ and $R_{DDA*-LAP}$. This shows that the performance ratios of DDA-LAP and DDA*-LAP algorithms are very good.

In addition to performance ratios, we have given the actual number of process steps on some datasets (case 1), as shown in Table 2.

**Table 2. Results of DDA, DDA* and other algorithms on simulated datasets. The numbers of process steps for every set are listed for different datasets.**

| Datasets | N | M*N | K | Length SH | Length LAP | Length Greedy-D | Length DDA-SH | Length DDA-LAP | Length DDA*-SH | Length DDA*-LAP |
|---|---|---|---|---|---|---|---|---|---|---|
| Case1 | 10,000 | 25,000 | 25 | 82 | 81 | 81,81,80 | 82,80,80 | 80,79,78 | 82,84,83 | 81,80,79 |
|  | 10,000 | 50,000 | 25 | 82 | 82 | 82,82,82, 81,80 | 83,82,80, 79,80 | 81,81,80, 74,77 | 84,80,82, 82,80 | 80,80,79, 81,79 |
|  | 10,000 | 25,000 | 35 | 98 | 97 | 100 | 98 | 97 | 100 | 97 |
|  | 20,000 | 25,000 | 25 | 82 | 81 | 81,80 | 82,80 | 81,78 | 82,82 | 79,80 |

It is obvious that the number of process steps by DDA and DDA* on multiple sets are shorter than those of Greedy-D, and shorter than those on single set (SH and LAP).

- **Results on Real Sequences**

For real gene sequences, we have selected EST sequences, plant sequences and rodent sequences from Genbank [14]. The $cost_{MM}$ cost function is used for analysis.

For each of the real gene sequences datasets with specific number of genes (G) and gene length (GL), we have selected a number of oligos (oligos per gene, OPG) of specific length (K). These oligos are used for analysis of performance of different algorithms. The same set of algorithms and setting are compared, and the lower bounds (LB) are also listed. The results are shown in Figure 6.

**Figure 6. Performance ratios of DDA, DDA* and other algorithms on real DNA datasets. X-axis shows the values of K, OPG, GL, G and N, respectively.**

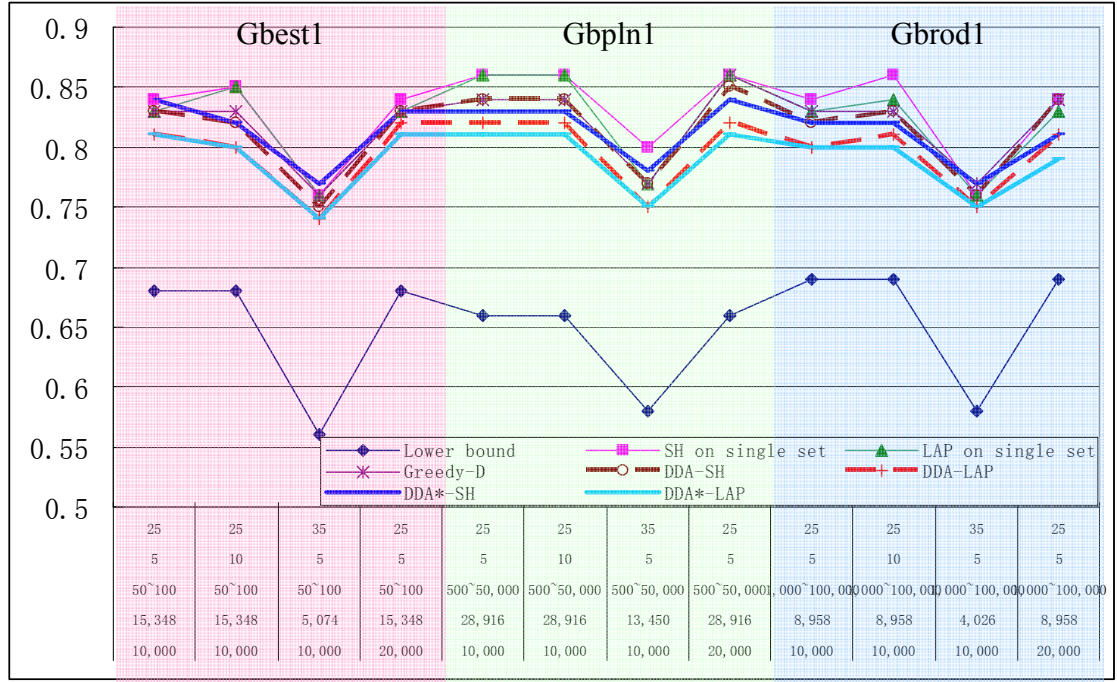

From results in Figure 6, the effectiveness of DDA and DDA* on real oligos is obvious. The results of $R_{Greedy-D}$ are always equal or bigger than $R_{DDA-SH}$ and $R_{DDA-LAP}$. $R_{DDA-SH}$(MSS) and $R_{DDA-LAP}$(MSS) are both about 0.01~0.05 less than the results of SH on single set, but only $R_{DDA-LAP}$(MSS) is smaller than he results of LAP on single set. These are more obvious for huge number of oligos (M*N≥1,000,000). Again, the performance ratio of DDA-LAP are better than DDA-SH, and DDA*-LAP is better than DDA*-SH. Also, the results are better on sets with smaller sizes, which is consistent with our previous analysis. And GC content still plays an important role. Compare with lower bounds, we observe that there is still about 0.1 to 0.2 gaps in performance ratios.

Comparing Figure 6 against Figure 5, we observe that performance ratios on real gene sequences are about 0.05 higher than those on simulated 4-alphabet sequences. This is probably due to the large variances in real gene sequences (and thus oligos).

For real text sequences, we have catagorized 19,997 sequences to 5 groups by contents. Here we have used the set size N=1,000, and set the sequences length (K) to be the length of each sequence. Note that for text sequences, the sequence lengths are not equal. The $cost_{SC}$ cost function is used for analysis. The results on text sequences are shown in Figure 7.

**Figure 7. Performance ratios of DDA, DDA* and other algorithms on real text sequences datasets. X-axis shows the values of K, M*N, N and dataset name, respectively.**

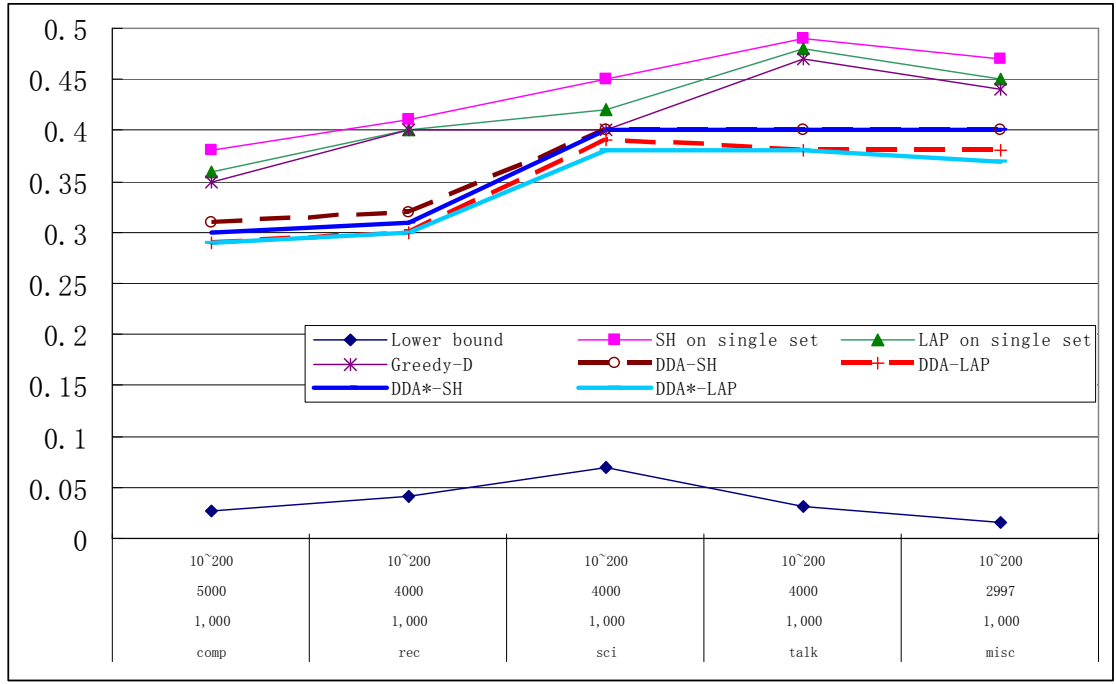

It is shown in Figure 7 that for text sequences, the performance ratios are generally low, mostly because of large alphabet size (the larger the alphabets size, the smaller the performances ratios). $R_{DDA-SH}$ and $R_{DDA-LAP}$, as well as $R_{DDA*-SH}$ and $R_{DDA*-LAP}$ are smaller than $R_{Greedy-D}$, $R_{SH}$ and $R_{LAP}$. These show that the DDA and DDA* are constantly better algorithms. Specifically, DDA-LAP outperforms DDA-SH on most of the datasets. We also note that the performance ratios of these algorithms are much larger than those of lower bounds. These are probably due to large alphabet for real text sequences, and variations among these text sequences are also large. This is especially significant on the results of "misc" dataset, which has max $|\Sigma|$. Moreover, based on large alphabet size, the lower bound also tends to be very loose.

We have also performed experiments on scheduling problems, but results are not shown in this paper. To generalize, the results on scheduling problems show that based on $cost_{SC}$ cost function, DDA*-LAP, DDA*-SH, DDA-LAP, DDA-SH and Greedy-D have increasing performance ratios on most of the datasets. DDA*-LAP, DDA*-SH, DDA-LAP and DDA-SH on multiple sets also outperform LAP and SH on single set (with all of the sequences in MSS).

- **Efficiency**

The time complexities of DDA and DDA* algorithms are already analyzed in previous section. In practice, the time for DDA and DDA* on PMSS problem is reasonable. Take large MSS for example, DDA-LAP takes around 10 minutes to process 50,000 sequences with K=25 and N=10,000, and DDA-SH takes around 30 seconds for the same sets. The DDA* algorithm is inherantly slow because of clustering. We observed that DDA*-SH takes around 20 minutes to deposit 50,000 sequences with K=25 and N=10,000, and DDA*-LAP takes around 30 minutes for the same sets. The majority of the time is spent on clustering, and the deposition process is very fast. Greedy-D algorithm is more efficient than DDA-SH and DDA-LAP, with about 10 seconds to process 50,000 sequences with K=25 and N=10,000.

The memory used by DDA and DDA* algorithms are also reasonable, with about 100MB for large scale sequence dataset (N=100,000, K=25 and M=10,000).

The programs of these algorithms are available upon request to the authors.

## 4. Discussions

Despite the increasing application of "multiple sequences sets" as a mathematical model, there is little research on the problem of efficient process of multiple sequences sets. In the design of process strategy for multiple sequences sets, the small number of process steps is desired such as in oligos synthesis, text processing and scheduling. For example, the smaller cost for the synthesis of multiple oligos is critical for the later chemical synthesis process (less time, less error etc.).

In this paper, we have defined the cost functions and performance ratios, based on which the problem of process multiple sequences sets (PMSS) is formulated as to minimize the cost of process. We have proposed greedy algorithms (Greedy-A and Greedy-D), the DDA and DDA* algorithm for solving the PMSS problem. Greedy algorithms are just generalization of greedy algorithms for single sequences set. Both DDA and DDA* first distribute sequences to different sets according to sequence features, and then process sequences on each set by SH algorithm or LAP algorithm. The difference between DDA and DDA* is that DDA relies on alphabet contents of the sequences to distribute sequences, while DDA* distribute sequences using clustering technique based on sequence profiles.

Experiments show that DDA and DDA* are effective for the PMSS problem: their costs and performance ratios are smaller than those of greedy algorithms (Greedy-A and Greedy-D). More specifically, the DDA* algorithm outperforms DDA algorithm in many instances. This indicates that distribution of sequences to multiple sets according to sequence features before processing sequences on each set is very beneficial. Additionally, we have shown that for the same set of sequences, distributing them to different set and then process then on respective sets is more cost efficient than processing these sequences on a single set. The DDA and DDA* algorithms are also efficient in time and space.

As the scale of applications based on multiple sequences sets is increasing, the PMSS problem will become more and more important. This work is our first attempt to solve the PMSS problem. We foresee a direct benefit of the algorithms that we have proposed is that the different synthesis strategies can be applied on different sets in parallel, and such parallel process will greatly reduce the time needed and facilitate the process of multiple sequences sets in microarrays synthesis, text processing, scheduling, as well as many other applications.

Multiple sequences sets model can be used for different applications; and specific applications may have specific requirements. For example, in the synthesis of microarrays, experimenters sometimes need only a selected number of oligos from a larger set of oligos pool to perform experiments. In these situations, how to select oligos, so that the later distribution and deposition process are more cost-effective is a more interesting yet complicated problem. The effective algorithms for these specific applications are also interesting, and this is another direction of our future work.

**Acknowledgement**

We thank anonymous reviewers for their thoughtful advices.**References:**

processing and scheduling. For example, the smaller cost for the synthesis of multiple oligos is critical for the later chemical synthesis process (less time, less error etc.).

In this paper, we have defined the cost functions and performance ratios, based on which the problem of process multiple sequences sets (PMSS) is formulated as to minimize the cost of process. We have proposed greedy algorithms (Greedy-A and Greedy-D), the DDA and DDA* algorithm for solving the PMSS problem. Greedy algorithms are just generalization of greedy algorithms for single sequences set. Both DDA and DDA* first distribute sequences to different sets according to sequence features, and then process sequences on each set by SH algorithm or LAP algorithm. The difference between DDA and DDA* is that DDA relies on alphabet contents of the sequences to distribute sequences, while DDA* distribute sequences using clustering technique based on sequence profiles.

Experiments show that DDA and DDA* are effective for the PMSS problem: their costs and performance ratios are smaller than those of greedy algorithms (Greedy-A and Greedy-D). More specifically, the DDA* algorithm outperforms DDA algorithm in many instances. This indicates that distribution of sequences to multiple sets according to sequence features before processing sequences on each set is very beneficial. Additionally, we have shown that for the same set of sequences, distributing them to different set and then process then on respective sets is more cost efficient than processing these sequences on a single set. The DDA and DDA* algorithms are also efficient in time and space.

As the scale of applications based on multiple sequences sets is increasing, the PMSS problem will become more and more important. This work is our first attempt to solve the PMSS problem. We foresee a direct benefit of the algorithms that we have proposed is that the different synthesis strategies can be applied on different sets in parallel, and such parallel process will greatly reduce the time needed and facilitate the process of multiple sequences sets in microarrays synthesis, text processing, scheduling, as well as many other applications.

Multiple sequences sets model can be used for different applications; and specific applications may have specific requirements. For example, in the synthesis of microarrays, experimenters sometimes need only a selected number of oligos from a larger set of oligos pool to perform experiments. In these situations, how to select oligos, so that the later distribution and deposition process are more cost-effective is a more interesting yet complicated problem. The effective algorithms for these specific applications are also interesting, and this is another direction of our future work.

**Acknowledgement**

We thank anonymous reviewers for their thoughtful advices.

**References:**

1. Pinedo, M.: Scheduling : theory, algorithms, and systems. Prentice Hall (2001)
2. Storer, J.A.: Data compression: methods and theory. Computer Science Press (1988)
3. Foulser, D.E., Li, M., Yang, Q.: Theory and algorithms for plan merging. Artificial Intelligence **57** (1992) 143 - 181
4. Sellis, T.K.: Multiple-query optimization. ACM Transactions on Database Systems (TODS) **13** (1988) 23 - 52
5. Cormen, T.H., Leiserson, C.E., Rivest, R.L., Stein, C.: Introduction to Algorithms. MIT Press and McGraw-Hill (2001)
6. Sankoff, D., Kruskal, J.: Time Warps, String Edits and Macromolecules: the Theory and Practice of Sequence Comparisons. Addison Wesley (1983)
7. Jiang, T., Li, M.: On the approximation of shortest common supersequences and longest common subsequences. SIAM Journal of Computing **24** (1995) 1122-1139
8. Barone, P., Bonizzoni, P., Vedova, G.D., Mauri, G.: An approximation algorithm for the shortest common supersequence problem: an experimental analysis. Symposium on Applied Computing, Proceedings of the 2001 ACM symposium on Applied computing (2001) 56 - 60
9. Timkovsky, V.G.: On the approximation of shortest common non-subsequences and supersequences. Technical report (1993)
10. Kasif, S., Weng, Z., Derti, A., Beigel, R., DeLisi, C.: A computational framework for optimal masking in the synthesis of oligonucleotide microarrays. Nucleic Acids Research **30** (2002) e106